\documentclass[a4paper,twoside]{article}

\usepackage{epsfig}
\usepackage{subcaption}
\usepackage{xcolor}
\usepackage{calc}
\usepackage{amssymb}
\usepackage{amstext}
\usepackage{amsmath}
\usepackage{amsthm}
\usepackage{multicol}
\usepackage{pslatex}
\usepackage{apalike}
\usepackage{algorithm}
\usepackage[noend]{algorithmic}
\usepackage{verbatim}
\usepackage{SCITEPRESS}     

\begin{document}

\title{Spatial K-anonymity: A Privacy-preserving Method for COVID-19 Related Geo-spatial Technologies}

\author{\authorname{Rohan Iyer\sup{1}\orcidAuthor{0000-0003-3055-9377}, Regina Rex\sup{1}\orcidAuthor{0000-0003-0327-8916}, Kevin P. McPherson\sup{1}\orcidAuthor{0000-0003-4977-8537}, Darshan Gandhi\sup{1}, Aryan Mahindra\sup{1}, Abhishek Singh\sup{1}\sup{2} and Ramesh Raskar\sup{1}\sup{2}}
\affiliation{\sup{1}PathCheck Foundation, Cambridge, USA}
\affiliation{\sup{2}MIT Media Lab, MIT, Cambridge, USA}
}


\abstract{There is a growing need for spatial privacy considerations in the many geo-spatial technologies that have been created as solutions for COVID-19-related issues. Although effective geo-spatial technologies have already been rolled out, most have significantly sacrificed privacy for utility. In this paper, we explore spatial k-anonymity, a privacy-preserving method that can address this unnecessary tradeoff by providing the best of both privacy and utility. After evaluating its past implications in geo-spatial use cases, we propose applications of spatial k-anonymity in the data sharing and managing of COVID-19 contact tracing technologies as well as heat maps showing a user's travel history. We then justify our propositions by comparing spatial k-anonymity with several other spatial privacy methods, including differential privacy, geo-indistinguishability, and manual consent based redaction. Our hope is to raise awareness of the ever-growing risks associated with spatial privacy and how they can be solved with Spatial K-anonymity.}

\onecolumn \maketitle \normalsize \setcounter{footnote}{0} \vfill

\section{\uppercase{Introduction}}
\label{sec:introduction}
\subsection{Motivation}
The COVID-19 pandemic has presented a multitude of challenges from public health to economic instability. Solutions for COVID-19 contact tracing and other purposes involve an exponential rise in the use of geo-location technologies and location-based services at every level of society. This has presented its own set of ethical challenges and adoption issues~\cite{adoptionprivacy1,hassandoust2020individuals}. Each individual's physical location and location history can be exploited by cybercriminals interested in mining personal information such as home, work, and school addresses as well as a daily itinerary, among others. The threat of leakage of information gets even more amplified in this scenarios since it also reflects the infection status of the person. Spatial privacy risks may likely extend beyond the individual, compromising families, coworkers, communities and even enterprises. Appropriate measures are necessary to ensure spatial privacy to be maintained in all solutions for the COVID-19 pandemic.

\subsection{Contribution}
This paper presents a developmental approach to addressing contact tracing spatial privacy concerns using Spatial K-anonymity \cite{Ghinita2010}. Spatial K-anonymity has been implemented in many pre-COVID-19 use cases \cite{Martin2015,sweeney2002,Allshouse2010,GkoulalasDivanis2010}; here, we discuss the unique applications that have emerged for spatial K-anonymity for the COVID-19 pandemic. We also compare the spatial K-anonymity approach with other existing solutions like Geo-indistinguishability and Consent Redaction to demonstrate the advantage of utilizing K-anonymity in COVID-19 contact tracing.

\section{\uppercase{Related Work}}
Before the COVID-19 outbreak, concerns about spatial data collection have been echoed worldwide regarding individual security and privacy. Due to these concerns, several policies, like the European Union's General Data Protection Regulation (GDPR) \cite{eu-269-2014}, outline identifying location data as personal, requiring other parties to anonymize the collected data or gain explicit consent from the user for spatial data collection. 

In light of current issues, spatial data plays an important role in surveillance, data sharing, and digital contact tracing as used by various government \cite{Park2020,Jacob2020,Jian2020} and corporate entities \cite{agpocct-20} in their efforts to combat COVID-19. 

Contact Tracing \cite{Lo2020} involves data collection to help public health officials better control and understand an infectious outbreak,  and it also informs individuals about their probable exposures to other infected people. Manual Contact Tracing (MCT) involves caseworkers working with infected persons to trace who they have been in close contact with during their infection period. MCT is laborious, and it relies on the infected person's ability to accurately recall people there were in close contact with \cite{Barrat2020}. It is also resource-intensive and not scalable; if many people are infected, each infected person would need to be assigned a caseworker. Digital Contact Tracing (DCT) is used to address the shortcomings of MCTs. It involves using technological devices such as smartphones and smartwatches to estimate close-range proximity It is more reliable than MCT, as it is more accurate in sensing people in close proximity \cite{Smieszek2016}. Despite the effectiveness and efficiency of DCT \cite{hbra2020}, there has been a lot privacy concerns regarding spatial data collection \cite{Rowe2020,Boulos2009}.

In the privacy domain, there have been efforts to ensure user spatial privacy rights are respected. State-of-the-art methods like k-anonymity \cite{GkoulalasDivanis2010}, l-diversity \cite{}, and t-closeness \cite{} are some of the few that are applied to various technologies that utilize spatial data. K anonymity assures that any query on data will give limited access and will maintain the anonymity of at least k individuals. This is usually achieved by using the concepts of purging or generalization. l-diversity covers for the cons of K anonymity by ensuring that there is significant attribute diversity between equivalence groups. However, it is difficult to accomplish with data having numerous Quasi Identifiers. t-closeness solves this problem by ensuring that the distribution of each sensitive attribute in an equivalence group is close to the distribution in the complete dataset. Thus, these methods help in maintaining user privacy immensely. 

Many spatial data protection measures have also been used to safeguard users' spatial privacy rights in various COVID-19 contact tracing applications. Some of the privacy measures applied as outlined in \cite{info:doi/10.2196/22594} involve pseudonymization or anonymization of spatial data, aggregation of geo-spatial and temporal data, data minimization, transparency, explicit consent, and ability to revoke consent by the users.

BlueTrace \cite{Bay2020BlueTraceAP} protects the users' privacy by using local storage for geo-location history and through the users' explicit consent to data location. They also provide the ability for a user to recount their given consent. Another contact tracing application, Epione \cite{trieu2020epione}, guarantees more robust privacy protection using a cryptographic protocol that securely delivers users' status to health authorities and ensures that users' contact information is not revealed to other non-desired parties. These are just some of the methods used to guarantee users' spatial data protection in digital contact tracing. A more unusual approach to this problem is TrackCovid \cite{Yasaka2020}, which avoids handling personal location data for contract tracing and uses anonymization via QR codes at specified checkpoints. \cite{chan2020pact} highlights possible privacy issues that can arise from the various methods of COVID-19 related contact tracing and suggests a Bluetooth pseudonymization approach to combat these privacy issues.

\section{COVID-19-based Applications of Spatial k-anonymity}

\subsection{Contact Tracing}

Location-based privacy is growing in popularity due to the data collection mechanisms of smartphones and other location-agnostic devices. During the COVID-19 pandemic, spatial privacy has proven to be of paramount concern to government officials, public health experts, and other medical professionals. In some countries, it is predicted that because of the consequences of sharing contact-tracing data, over 70\% of patients who have contracted COVID-19 were at risk of being identified by location, gender and/or age \cite{Jung2020}. 

GPS provides an advantageous solution to contact tracing than other solutions, given other technologies, such as outbreak response tools or symptom tracking tools. Particularly, GPS based solutions can transmit data in background mode via smartphones and wearable devices, provide public space and infection proximity data, and have cross-platform functionality inherent to the data \cite{}. When digital contact tracing employs a GPS-based solution for tracing individuals, it often queries highly specific geo-location data of a user. This data is quite sensitive as it tracks the real time location of a user and stores it onto a database that could be potentially used for tracking and tracing purposes. Spatial k-anonymity employs certain methodologies that convert highly accurate geo-location data into a more granular form while preserving the effectiveness of location-based contact tracing. 

Imagine person X queries location-based data from address Y. An attacker may use this information to associate person X’s real-time location with high accuracy.  Spatial k-anonymity solves this problem by employing methods that ensure user X’s location-based query has a probability larger than 1/k, where k is a defined anonymity requirement.  

A widely used method to calculate k involves masking a user's location by deploying specific geomasking techniques and relocating the user to a new buffer location. Geomasking is a general category of methods that is used to mask or change the geographical location of a user or patient in an unpredictable way that maintains that user or patient's relationship to a particular geo-location and their status as it concerns a disease infection or outbreak \cite{}. The k-anonymity value is the number of all potential residential locations inside this buffer area, including the original user location. The value of 1/k is thus a measure of the person's disclosure risk in question (the probability of being re-identified) after geomasking. The smaller the 1/k, the more difficult it is to re-identify the person's real location.

\subsection{Heat Map}

The incorporation of heatmaps is primarily for better visualization and understanding of the travel history of the user over a period of time.  With the help of the heatmap feature the user can realise the travel history and make optimum decisions while planning their next travel journey. 

\begin{figure*}[t!]
  \centering
   {\epsfig{file = 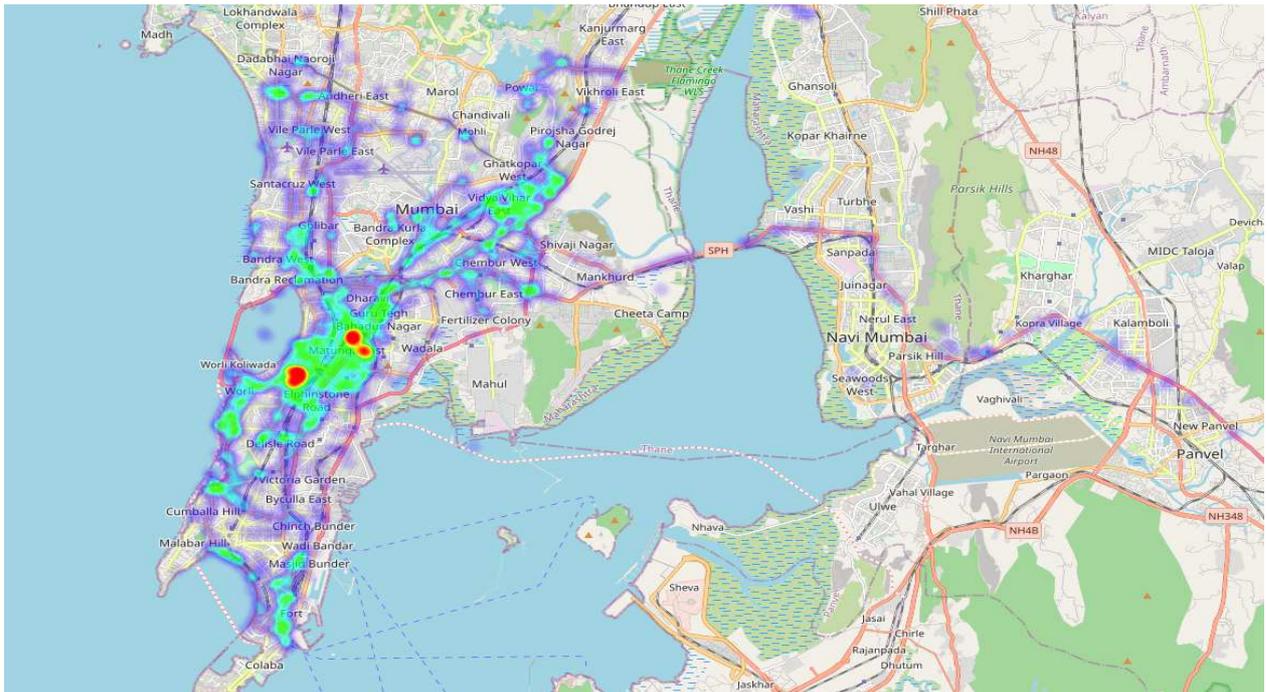, width = 16.51cm,
   height = 9.144cm}}
  \caption{An example of a user's location history represented as a heat map (Red-orange $\rightarrow$ heavily visited, Green $\rightarrow$  regularly visited, Blue $\rightarrow$  least visited).}
  \label{fig:heatmap-example}
  \vspace{-0.1cm}
\end{figure*}

Case 1: Restricted to an individual user 

In this case the user can get detailed information about their travel history and they can use this information to further make their travel plans. 

The user can map a circle or any arbitrary shape to distinguish the regions as a public or private one. Based on the segregation, the user can be shown only their travel history in the public region. This would help them to understand their travelling trends. Also, there can be potentially marked several points of danger and caution on the map and accordingly notify the user if they have crossed paths during their journey. A track of the last 7-14 days should be kept ideally since that would help to keep track of the exposure period. 

The primary objective of this is to keep the user aware of their surroundings and help them make concrete decisions while making their next travel plan.The first option should be the most suited and feasible as well.

Case 2: Combination of all the users data 

In this case, the methodology is to club the details of all the users around a specific user and with the help of that a combined heat map can be plotted. This would potentially help the user get a good understanding of their surroundings. Again, classification of zones can be implemented here with the RED region as a danger zone and alert the user if they are travelling by that particular area. 

Many of the new metrics are based on uncertainty. In this paper here the authors suggested a test how easily and quickly the location of the user can be traced and identified from the anonymous collection of users. It was observed that the major focus was on the path  a user might have taken. 
Thus, one of the key parameters metric to track would be to identify the size of the cluster, larger the size, larger the ambiguity and more difficult to track the individual user. 

A few suggestions would be to showcase the exact coordinates as street or even the town/ city coordinates, this would add a high amount of ambiguity in the process of identifying the exact location of the user thus helping them stay protected and preserve their privacy. 

\section{Comparing Spatial K-anonymity With Other Privacy-Preserving Methods}

In this section we will give a detailed description of the various spatial privacy approaches: K-anonymity, geo-indistinguishability, and consent redaction. We will also show how these approaches differ from each other and spatial k-anonymity, highlighting their strengths and weaknesses in digital contact tracing and other COVID-19 related applications.

K-anonymity \cite{GkoulalasDivanis2010} is a well known anonymization technique for spatial privacy. This technique refers to the process of transforming a data set that contains personal or identifying information to a dataset that strips away any information that could identify an individual, making it hard for other parties to determine the identity of an individual from the transformed data set. In this paper, we will be focusing on the K-Anonymity algorithm for protecting spatial privacy. K-anonymity is characterized as the number of individuals, which is a distinct value ‘k’,  associated with alike attributes like gender, city, etc such that one particular individual cannot be identified from the data set. Since the dataset consists of geographical identifiers as well, the traditional k-anonymity algorithm can be expanded to spatial k-anonymity. Spatial k-anonymity is a widely used method to evaluate the degree of geoprivacy obtained following the implementation of a particular geo masking technique.

Spatial k-anonymity \cite{} works the best for conventional geo-spatial datasets wherein only one geographical area is identified for each individual, however, it has important shortcomings when dealing with individual GPS datasets. Spatial K-anonymity algorithm is highly regarded since rather than having a single geographic location that can disclose a person's identity, it provided multiple several everyday activity locations, for example, home, offices, and shopping venues that can be easily identified from a GPS dataset, which makes it difficult to zero down a single person. Having data of a single location associated with a person can make an individual vulnerable by providing a wealth of geographic data to hackers and can be utilized for similar malpractices. However, since most people spend only a small amount of time at home and spend a significant amount of time away from home, having multiple locations increases the scope of user privacy. It is important to consider the disclosure risk associated with all daily activity locations in a detailed manner.

Geo-indistinguishability \cite{andres-geo-indistinguishability-2013} is a metric used for location-based systems that protect the user's exact location while allowing approximate information, typically needed to obtain a specific desired service, to be released. The underlying motivation for geo-indistinguishability comes from differential privacy but more handcrafted for the spatial data. It is focused on protecting the user's location within a radius, , with a level of privacy that depends on  that corresponds to a generalized version of the well-known concept of differential privacy. The  value must be optimized in consideration of both privacy and utility, and it must be large enough to fit the notion of privacy. That is, an attacker with potential auxiliary information should not be able to pinpoint the user's identity. For example,  must also be small enough to allow digital contact tracing to accurately identify users who have been in contact with a user who has tested positive. Like spatial K-anonymity, geo-indistinguishability guarantees spatial privacy needs for location-based systems such as digital contact tracing apps. \cite{dams2020} uses differential privacy to release a sketching data structure of location data. Aside from this, there have not been any application of geo-indistinguishability in digital contact tracing.

Manual Consent Redaction is a privacy-preserving methodology that enables users to revoke given consent to data collection for specific locations or all locations tracked by geo-spatial technologies. This technique offers users greater control over their data, as they can choose to opt-in or out of the service or select what data points to exclude from external servers. In manual consent redaction, the user can choose to revoke their consent for geo-spatial technologies to access all their locations or just for specific locations. If the user desires the latter, they must manually select which locations to exclude from the data collection process, and these would also be removed from the external servers. Choosing the former means the user can not use the geo-spatial technology, and all their data should be removed from the servers. This approach gives user's control over their privacy and is a highly interpretable notion of privacy from a user standpoint. Its disadvantages includes the risk of human error and it has less privacy-utility tradeoff.

Most contact tracing apps only give the users the option of fully consenting to data collection or revoking their consent. Using this approach, the user can only use the contact tracing app if they agree to data collection. Otherwise, they can not specify if they want only specific locations redacted. Unlike the other privacy-preserving methods above, manual consent redaction alone does not fully guarantee spatial privacy. It is more restrictive, as revoking consent bars a user from using the technology. For example, \cite{Bay2020BlueTraceAP} uses consent revocation in combination with other privacy-preserving methods to ensure users' privacy. The more user-friendly approach, where users can exclude specific data points if they choose to, does not guarantee the remaining data collected are non-identifying. i.e., attackers can still identify users based on their DAL even if they redact private locations from data collection. 

\section{Developmental Solution}

The idea of a spatial k-anonymity development solution is that it takes a  list of user’s locations (L), and creates a bound portion of anonymized geographical locations (anonymized-L) that is shared with at least K-1 other users. 

Pulling heavily on the previous work of \cite{Cassa2006}, which addresses the issues of using an anonymizing algorithm while preserving the fidelity of containing an outbreak through epidemiology data, the features of our algorithm are as follows:

\begin{enumerate}
    \item \textbf{Population Density-Based Gaussian Spatial Skew.} We hope to “blur” any location the user deems as private. As such, we need to take the user’s addresses that are private (represented in longitude and latitude), and skew using a random offset based on Gaussian distribution which has a standard deviation that is inversely proportional to the area’s population density. This allows for smaller transposition of people in high density areas while maintaining k-anonymity. 
    \begin{itemize}
        \item \textbf{Individual point Gaussian Blur}. Points will have to be skewed a minimum distance. To do this, we develop two randomly selected variables, $\sigma _{x}$ and $\sigma _{y}$ which represent standard deviations from a normal distribution to give the distance and direction of a user’s displacement. From here, two displacement values ($(d_{x}, d_{x})$).  Gaussian blur, a concept used in image processing, takes a float or integer representation and applies noise in line with a Gaussian distribution. 
    \end{itemize}  
    
    \item \textbf{Combining with the Anonymization Algorithm of Cassa et al.} Assuming we have some context data like US Census or other state-sponsored data collection, we want anonymization skewed inversely with population density. That is, if someone lives in a highly-dense area, they should move a shorter distance in the anonymized mode, whereas rural participants move larger distances in the anonymized mode. 
    
    \item \textbf{Anonymization Multipliers.} In order to control for population and age-based density issues, multipliers from the Cassa et al. paper will be employed. They are as follows:
    \linebreak
    APDM = Age-based population density multiplier
    
    TPDM = Total population density multiplier
    
    AGPD = Average age group population density
    
    UBGAD = User's block group age density
    
    ATPD = Average total population density
    
    UBGPD = User's block group population density
    
    AM = Age Multiplier
    
    APD = Age-based population density
    
    CM = Combined multiplier
    
    \vspace{3mm}
    \begin{equation}\label{eq1}
       APDM = AGPD/UBGAD
    \end{equation}
    \vspace{3mm}
    \begin{equation}\label{eq2}
        TPDM = ATPD/UBGPD
    \end{equation}
    
    Equations (1) and (2) combine to form a combined multiplier that looks like this:

    \begin{equation}\label{eq3}
        CM = AM * APD + (1 - AM) * TPDM
    \end{equation}
    
    The overall Anonymization Multiplier is as follows, with c being a scaling factor that can alter the overall skew of a coordinate pair:
    
    \begin{equation}\label{eq4}
        CM = AM * APDM + (1 - AM) * TPDM
    \end{equation}
\end{enumerate}


These solutions should prove to hold up to traditional spatial-k anonymity privacy standards that any nefarious effort to identify the user’s location can be pinpointed with no greater than 1/K probability \cite{Ghinita2010}.

\section{Conclusions}

Currently, we have built an application to demonstrate the concept of spatial privacy protection of the user. For the same we managed to build a working front end and back end product. In future, we plan to upscale this and make it production ready to be deployed for real life use cases. For the same we plan to shift from using SQLite as our database to others such as Postgres for handling heavy load. Also, we plan to add several new features such as heatmap generation for the user to help them understand their travel patterns, mark points/places of heavy concentration of users and declare it to be a zone of high risk, also identify and mark the points which are public places. All these features would help the user to plan their travel journey better and also help understand the surroundings they reside in a better way.

\section*{\uppercase{Acknowledgements}}

Special thanks to PathCheck Foundation for supporting our paper, including Ramesh Raskar, Aryan Mahindra, and Haris Nazir.

\bibliographystyle{apalike}
{\small
\bibliography{main}}


\end{document}